\documentclass[10pt,onecolumn]{IEEEtran}
\usepackage{amsmath,amssymb}
\usepackage{algorithm,algpseudocode}
\usepackage{fancybox}
\usepackage{tikz}
\usetikzlibrary{arrows, decorations.markings}
\usepackage{pdfpages}

\tikzstyle{vecArrow} = [thick, decoration={markings,mark=at position
   1 with {\arrow[semithick]{open triangle 60}}},
   double distance=1.4pt, shorten >= 5.5pt,
   preaction = {decorate},
   postaction = {draw,line width=1.4pt, white,shorten >= 4.5pt}]
\tikzstyle{innerWhite} = [semithick, white,line width=1.4pt, shorten >= 4.5pt]

\tikzstyle{vecArrow} = [thick, decoration={markings,mark=at position
   1 with {\arrow[semithick]{open triangle 60}}},
   double distance=1.4pt, shorten >= 5.5pt,
   preaction = {decorate},
   postaction = {draw,line width=1.4pt, white,shorten >= 4.5pt}]

\newtheorem{theorem}{\hspace*{1pc}Theorem}
\newtheorem{corollary}{\hspace*{1pc}Corollary}
\newtheorem{lemma}{\hspace*{1pc}Lemma}

\begin{document}

\bibliographystyle{unsrt}

\setlength{\parindent}{1pc}

\title{Loss Tomography in General Topology}
\author{Weiping~Zhu \thanks{Weiping Zhu is with University of New South Wales, Australia}}
\date{}
\maketitle

\begin{abstract}
Although there are a few works reported in the literature considering loss tomography in the general topology, there is few well established result since all of them rely either on simulations or on experiments that have many random factors affecting the outcome. To improve the situation, we address a number of issues in this paper that include a maximum likelihood estimator (MLE) for the general topology, the statistical properties of the MLE,  the statistical properties of a frequently referred estimator called the moving variance and weighted average (MVWA), and a renewed MVWA that removes the restriction of knowing variance in advance from the MVWA. The statistical properties covers  minimum-variance unbiasedness, efficiency, and variances of the estimates obtained by the estimators. Given the properties, we can evaluate the estimators without the need of simulations. To verify the properties, a simulation study is conducted that confirms the accuracy of the findings.
\end{abstract}


\section{Introduction}\label{section1}

Network characteristics, such as link-level loss rate, delay distribution, available
bandwidth, etc. are valuable information to network
operations, development and researches. Because of this,
 a considerable attention has been given to network
measurement, in particular to the networks that do not support direct measurement. To overcome the restrictions, network tomography was
proposed in \cite{YV96}, where the author suggests the use of
end-to-end measurement and statistical inference to accomplish the task.
Many works have been reported since then that can be divided into a number of areas, such as loss tomography \cite{BDPT02}, delay tomography
\cite{LY03,TCN03,PDHT02, SH03, LGN06},
and loss pattern tomography \cite{ADV07}, etc.,  naming after the characteristic of interest. Despite this, some fundamental issues have not been addressed or answered in loss tomography, especially the statistical properties of the estimators proposed previously. This paper aims at addressing some of the issues in loss tomography for the general topology that include a MLE, a renewed MVWA, the statistical properties of them.

 Although a large number of estimators have been proposed for the networks of the tree topology  \cite{CDHT99, CDMT99, CDMT99a, CN00, XGN06, BDPT02,ADV07, DHPT06, ZG05, GW03}, there have been only a few attempts for the general topology.  Among the few attempts, \cite{BDPT02}  is  the  most cited one that uses a number of multicast trees to cover a general network and uses the estimator proposed in \cite{CDHT99} for the tree topology to estimate the loss rates of the links.
If a link covered by a number of multicast trees, the same number of estimates are obtained for the link that can be different from each other.  To resolve the differences among the estimates, a weighted average scheme called the moving variance and weighted average (MVWA) is proposed in \cite{BDPT02} that aims at minimizing the variance of the estimate obtained by a weighted average method.
The expectation-maximization (EM), the MVWA, and other alternatives are compared by simulations because of a lack of statistical analyses for the estimators. However,  some of the results received from simulations or experiments are contradictory to each other. For example, the criticism of the MVWA in \cite{BDPT02, DLZL15} are not  justified since there is a lack of evidence  in theory to support the criticism. We argue  in this paper that although there are some practical issues on the way of implementing the MVWA, the philosophy of the MVWA is sound and can be used as a reference to evaluate other estimators.
To correctly evaluate an estimator, we need to have the statistical properties of the estimator, such as unbiasedness, efficiency, variance. Unfortunately, there has been few attempt in this direction.

Apart from statistical properties,  we, in this paper, propose an estimator for the general topology that has three components: a decomposition principle, two estimators, and an estimation order, that together form a MLE.  The statistical properties of the proposed MLE are derived that shed light on how to create an efficient MVWA that eventually leads to a number of alternatives. The contribution of this paper to loss tomography is articulated in the following aspects:

\begin{enumerate}
\item A
divide-and-conquer strategy is proposed
to divide a general network
into a number of zones, called d-trees, that have consistent observation, i.e. all nodes in a d-tree observe the probes multicasted by the same source(s).
\item A MLE is
proposed to estimat the pass rate, defined as the success rate of a probe passes a network segment that can be a link/path/tree, of a d-tree that receives probes from a number of sources and the statistical properties of the MLE are derived, including  minimum-variance unbiasedness, efficiency, and variance.
\item Apart from the MLE, another estimator is proposed to estimate the loss rate of an upstream link of a d-tree. The statistical properties of the estimates obtained by the estimator are obtained.
\item The low bound of the variance of the estimates obtained by the MVWA is derived  that is identical to the variance of the estimates obtained by the MLE. Based on the low bound, a number of alternatives are proposed on the basis of moving average.
\item The proposed estimators are compared with that proposed for the tree topology and the alternatives proposed on the basis of the MVWA in simulations to verify the statistical properties identified above.
\end{enumerate}

The rest of the paper is organized as follows. In Section \ref{section2}, we
present the related works and point out the direction to find the MLE for the general topology. We then introduce the assumptions and notations used in this paper  in
Section \ref{section3}. The likelihood function connecting observations to the parameter to be estimated is created and the sufficient statistic is identified  in the section.
In Section \ref{section4}, we propose two estimators, one is for the intersection of a number of multicast trees and the other for the upstream links of an intersection. The
statistical properties of the proposed estimators are derived and presented in Section \ref{section5} and the alternatives of the MVWA are proposed in the section.
In Section \ref{section8} we present the result of a simulation study to verify the statistical properties identified in Section \ref{section5}.
 The last section is devoted to
concluding remark.

\section{Related Works, Problems and Direction}
\label{section2}

\subsection{Related Works}
Multicast Inference of Network Characters (MINC) is the pioneer of putting the ideas proposed in \cite{YV96} into practice, where a
Bernoulli model is used to model  the behaviors of a probe passing a link/path. Using
this model, the authors of \cite{CDHT99} propose an estimator for the tree topology to estimate the pass
rate of a path connecting the source to a node. The estimator is expressed
in a polynomial that is one degree less than the number of descendants
of the node \cite{CDHT99, CDMT99, CDMT99a}.  To ease the
concern of using numeric method to solve a higher degree polynomial $(
> 5 )$, the authors of \cite{DHPT06} propose an explicit estimator and claim the estimator has the same asymptotic variance as that obtained by the polynomial to first
order. Unfortunately, this claim is not correct \cite{Zhu15}.

Considering the unavailability of multicast in some networks, Harfoush
{\it et al.} and Coates {\it et al.} independently proposed an alternative that uses a number of unicast packets to mimic a multicast \cite{HBB00},
\cite{CN00}.  Coates {\it et al.} also suggested the use of the
EM algorithm to estimate link-level loss rates. In addition,
Rabbat {\it et al.} in \cite{RNC04} consider network tomography on
general networks and find a general network comprised of multiple
sources and multiple receivers can be decomposed into a number of 2 by
2 components. The authors further proposed the use of the generalized
likelihood ratio test to identify network topology. To improve the
scalability of an estimator, Zhu and Geng propose a bottom up
estimator  for the tree topology in \cite{ZG04}, which is
found to be topology independent in \cite{ZG05}. The estimator adopts a
step by step approach to estimate the loss rate of a link, one at a
time from bottom up. Despite the effectiveness,
scalability, and extensibility to the general topology,
 the estimate obtained by the estimator is not the MLE because the statistics used
 by the two estimators are not sufficient ones.

Apart from the estimators proposed for the tree topology, there have been a number of attempts to extend loss tomography from the tree topology to the general one.  The work reported in \cite{BDPT02} is such an attempt that considers to use a number of multicast trees to cover a general network and considers each of the multicast trees independently in order to use the the estimator proposed in \cite{CDHT99} to estimate the loss rate of a link. As a result, a number of estimates are obtained for the links shared by a number of multicast trees that are different from one another because each of them uses a part of the observation only. In order to have an estimate for a shared link, the MVWA is proposed that aims at minimizing the variance of the estimate. Although the objective is achievable in theory,
in practice there are a number of issues that need to be addressed and the most important two are how to obtain the variances for computing the weight of an estimate and how to ensure the accuracy of the variances.
 In addition, the EM algorithm is introduced  to replace the MVWA in \cite{BDPT02} because of the poor performance of the MVWA in experiments. Despite this, it is not clear whether the estimate obtained the EM algorithm is the maximum likelihood estimate since there is no likelihood equation. Recently,  \cite{DLZL15} shows the EM algorithm used in \cite{BDPT02} has its own weaknesses in terms of computation complexity. Then,  an revised EM algorithm is proposed to replace the origin \cite{DLZL15}. Unfortunately, almost all of the findings are not conclusive since  most of them are gained from simulations and experiments that have many random factors affecting the outcome.

\subsection{Problems and Direction}

The fundamental difference between a network of the general topology and that of the tree one is whether a node can have more than one parents. If a number of multicast trees are used to cover a general network and the probes are multicasted from the sources to  receivers, the sources sending probes to the receivers may not be consistent to the receivers. For example, two receivers attached to a network may observes the probes sent by different sources. Although it is possible to write a likelihood function for this scenario, it is not easy to derive a likelihood equation from the likelihood function, in particular if we intend to use all available information in estimation. To solve the problem, we,
rather than dividing a general network into a number of multiast trees according to the sources as that reported in  \cite{BDPT02}, need to use a different way to divide
 a general network into a number zones to eliminate inconsistent observation in a zone. To achieve this, a general network is divided  on the basis of the coverage of the multicast trees, each component is called a d-tree, in which each node has a consistent observation. For instance, the network shown in Fig. \ref{multisource} can be divided into 3 d-trees. All of the d-trees obtained from such a decomposition are either tree or multicast tree  in terms of shapes, where the former is at the bottom of the general network, formed by the intersections of the multicast trees covering the network, and the latter at the top of the intersections.  Because of this, there is a dependency between the estimates obtained in two neighbour zones that one is at the top of the other. To support such a decomposition, we need two estimators, one for a shape of the zones. The estimator for the  shape of multicast tree needs to take into account the estimate obtained from its downstream neighbour.

\section{Assumption, Notation, and Statistics}
\label{section3}

\subsection{Assumption}\label{section3.a}
A similar assumption as that presented in \cite{YV96} and \cite{CDHT99} is used in this paper, which assumes the probes sent in an experiment are
independent from each other, the losses occurred on the links are  temporal and spatial  independent, the path from a source to a receiver remains constant during the experiment, and the traffic of the network remains statistically stable.  Then, the observation obtained in probing is considered independent identical distributed
($i.i.d.$) and the likelihood function is equal to the product of the individual observations. In addition, although there may have various shapes created by the coverage of the multicast trees for a general network, we in this paper only consider two stated above since all others can be converted to the two.

\subsection{Notation}
To make the following discussion rigorous, a large number of symbols are used  in the following discussion that may overwhelm some of the readers who are not familiar with loss tomography. To assist the readers, the symbols are gradually introduced through the paper, while the frequently used symbols are introduced here and the others will be brought up later when needed. In addition,  the most frequently used symbols and their meanings are presented in Table \ref{Frequently used symbols and description} for quick reference.

 Let ${\cal N}=\{T^1, T^2,\cdot\cdot, T^k\}$ be the set of multicast trees used to cover the general
network of interest, where $T^i$ is the multicast tree having source $i$ attached to its root. Let $S=\{1, 2, \cdot\cdot,
k\}$ represent all of the sources attached to ${\cal N}$. $T^i$ and $T^j, i, j \in S$ and $i \neq j $, may be partially overlapped, where the  largest overlap between them is called the {\it
intersection} of them. Each intersection is a tree, not a multicast tree,  and the root of an intersection is called
the {\it joint node} that has more than one parents. Let $J$ denote all of the joint
nodes in ${\cal N}$ and $S(i), S(i) \subset S,  i \in J $ denote all of the sources that send probes to the intersection rooted at node $i$, called intersection $i$ later. In addition, we use $V$ and
$E$ to denote the nodes and links in ${\cal N}$, respectively. Each  member of $V$ and
$E$ has  a
unique number. Apart from the above, $|V|$ and $|E|$ are used to denote the number of
nodes and the number of links in $V$ and $E$, respectively. In addition,  $R$ is used to denote all of the receivers attached to ${\cal N}$ and $R(i)$ is used to denote the receivers attached to the subtrees rooted at node $i$.

In the tree topology, we normally make such an arrangement that  link $i$ points to node $i$ since it makes both the notation and the discussion simpler. Nevertheless, such an arrangement is no longer possible  in the general topology since a node may have multiple parents. If we use $T(i), i \in V$ to denote the subtree rooted at node $i$ and $TL(i), i \in E$ to  denotes the multicast subtree that has link $i$ as its
root link, we cannot conclude that $\forall i, T(i) \subset TL(i)$. Similarly, if we use  $RL(i)$ to denote the receivers attached to $TL(i)$ and $R(i)$ to denote the receivers attached to $T(i)$, we no longer have $\forall i, R(i)= RL(i)$ since there is no longer such a correspondence in the general topology.

Since a node  in ${\cal N}$ can have a number of descendants and a number of  parents, we use $d_i$ and $f_i$ to denote
the descendants attached to node $i$ and the parents of node $i$, respectively. In addition, $f_1^s(i),  f_1^s(i) \in f_i$, simply $f^s(i)$,
denotes the parent of node $i$ located on the path from node $i$ upward to $s$ and  $f_l^s(i)$ is used
to denote the ancestor that is $l$ hops away from node $i$ upward to source $s$.
Further, let $a(s,i)=\{f^s(i), f_2^s(i),\cdot\cdot, f_k^s(i)\}$, where
$f_{k+1}^s(i)=s$, denote the ancestors of node $i$ on the path upward to $s$.
If there are multiple parents for node $i$, we use  $a(i)=\{a(s,i), s \in S(i)\}$ to denote all of the ancestors of node $i$.

To use active approach to estimate the loss rates/pass rate, we need to send probes from $S$ to $R$, where the probes sent by $s, s \in S$ target to a group receivers. Let $n^s, s \in S$, be the
number of probes sent by source $s$ to the receivers attached to $T^s$.
Each probe sent by $s$, say $o$, gives
rise of an independent realisation $X^s(o)$ of the passing process $X^s$ at the links transmitting the probe, which is also recorded at the nodes that transmits the probe unless the probe either reaches a leaf node or is lost.
Let  $Y_k^s(o), k \in V, s \in S$ denote the state of node $k$ for probe $o$ sent by source $s$. Since we only know the states of the nodes in $R$, the  state of an internal node, say $k$, for a probe, say $o$, can only be inferred from the observation of the receivers attached to $R(k)$, where we have $Y_k^s(o)=1$ if $\exists z, Y_z^s(o)=1, z \in R(k)$, $Y_k^s(o)=0$ otherwise.
Let $\Omega_s=\{(Y_k^s(o))|k \in R; o=1,2,...,n^s\}$ be the observation obtained by the receivers attached to $T^s$ for the probes sent by $s$ that can be divided into a number of segments according to subtrees, where $\Omega_s(i)=\{(Y_k^s(o))|k \in R(i); o=1,2,...,n^s\}$ denotes the observation of $R(i)$ for the probes sent by source $s$. If $T(i) \not\subset T^s$, $\Omega_s(i)=\emptyset$. Further, let $\Omega=\{\Omega_s|s \in S\}$ denote the observation obtained in an experiment.
\begin{table}[htdp]
\caption{Frequently used symbols and description}
\begin{center}
\begin{tabular}{|c|l|} \hline
Symbol & Desciption \\\hline
$T^i$ & multicast tree $i$ \\ \hline
$T(i)$ & the subtree rooted at node $i$ \\ \hline
$TL(i)$ & the multicast subtree rooted at link $i$ \\ \hline
$d_i $& the descendants attached to node $i$ \\\hline
$R(i)$ & the receivers attached to $T(i)$ \\ \hline
$RL(i)$ & the receivers attached to $TL(i)$ \\ \hline
$A(s,i)$ & the pass rate of the path from s to node $i$ \\ \hline
$\beta_i$& the pass rate of $T(i)$ \\ \hline
$Y_k^s(o)$ & the state of node $k$ for probe $o$ sent by $s$ \\ \hline
$\gamma_i(s)$ & the pass rate from source $s$ to $R(i)$ \\ \hline
\end{tabular}
\end{center}
\label{Frequently used symbols and description}
\end{table}%



\begin{figure}
\begin{center}
\begin{tikzpicture}[scale=0.2,every path/.style={>=latex},every node/.style={draw,circle,scale=0.6}]
  \node            (a) at (15,30)  { 16 };
  \node            (b) at (25,30)  { 0 };
  \node            (c) at (15,24) { 17 };
  \node            (d) at (25,24) { 1 };
 \node            (e) at (10,18)  { 18 };
  \node            (f) at (20,18)  { 2 };
  \node            (g) at (30,18) { 3 };
  \node            (h) at (6,12) { 19 };
  \node            (i) at (13,12)  { 20 };
  \node            (j) at (18,12)  { 4 };
  \node            (k) at (23,12) { 5 };
  \node            (l) at (28,12) { 6 };
  \node            (m) at (34,12)  { 7 };
  \node            (n) at (4,6)  { 21 };
  \node            (o) at (8,6) { 22 };
  \node            (p) at (11,6) { 23 };
  \node            (q) at (14,6)  { 24 };
  \node            (r) at (16.5,6)  { 8 };
  \node            (s) at (19,6) { 9 };
  \node            (t) at (21.7,6) { 10 };
   \node            (u) at (24.5,6)  { 11 };
  \node            (v) at (27.5,6)  { 12 };
  \node            (w) at (30.2,6) { 13 };
  \node            (x) at (33,6) { 14 };
  \node            (y) at (36,6) { 15 };

  \draw[->] (a) edge (c);
  \draw[->] (b) edge (d);
  \draw[->] (c) edge (e);
  \draw[->] (c) edge (f);
  \draw[->] (d) edge (f);
  \draw[->] (d) edge (g);
  \draw[->] (e) edge (h);
  \draw[->] (e) edge (i);
  \draw[->] (f) edge (j);
  \draw[->] (f) edge (k);
  \draw[->] (g) edge (l);
  \draw[->] (g) edge (m);
  \draw[->] (h) edge (n);
  \draw[->] (h) edge (o);
  \draw[->] (i) edge (p);
  \draw[->] (i) edge (q);
  \draw[->] (j) edge (r);
  \draw[->] (j) edge (s);
  \draw[->] (k) edge (t);
  \draw[->] (k) edge (u);
  \draw[->] (l) edge (v);
  \draw[->] (l) edge (w);
  \draw[->] (m) edge (x);
  \draw[->] (m) edge (y);
\end{tikzpicture}
\caption{A Network with Multiple Sources} \label{multisource}
\end{center}
\end{figure}
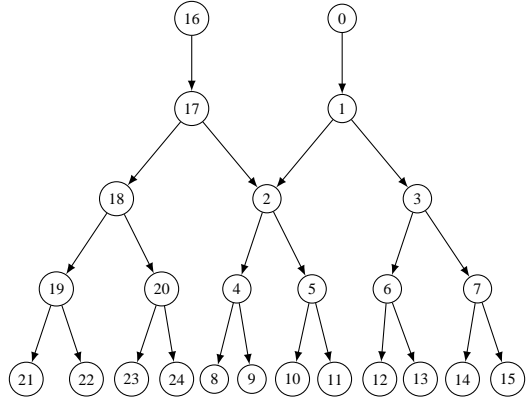

\subsection{D-tree and Dependency}

\begin{figure}
\begin{center}
\begin{tikzpicture}[scale=0.2,every path/.style={>=latex},every node/.style={draw,circle,scale=0.5}]
  \node            (a) at (13,30)  { 16 };
 \node            (a1) at (17,25)  { 16 };
  \node            (b) at (27,30)  { 0 };
   \node            (b1) at (23,25)  { 0 };
  \node            (c) at (13,24) { 17 };
  \node            (d) at (27,24) { 1 };
 \node            (e) at (10,18)  { 18 };
  \node            (f) at (20,17)  { 2 };
  \node            (f1) at (16,20)  { 2 };
  \node            (f2) at (25,20) {2};
  \node            (g) at (30,18) { 3 };
  \node            (h) at (6,12) { 19 };
  \node            (i) at (13,12)  { 20 };
  \node            (j) at (18,12)  { 4 };
  \node            (k) at (23,12) { 5 };
  \node            (l) at (28,12) { 6 };
  \node            (m) at (34,12)  { 7 };
  \node            (n) at (4,6)  { 21 };
  \node            (o) at (8,6) { 22 };
  \node            (p) at (11,6) { 23 };
  \node            (q) at (14,6)  { 24 };
  \node            (r) at (16.5,6)  { 8 };
  \node            (s) at (19,6) { 9 };
  \node            (t) at (21.7,6) { 10 };
   \node            (u) at (24.5,6)  { 11 };
  \node            (v) at (27.5,6)  { 12 };
  \node            (w) at (30.2,6) { 13 };
  \node            (x) at (33,6) { 14 };
  \node            (y) at (36,6) { 15 };

  \draw[->] (a) edge (c);
  \draw[->] (a1) edge (f);
  \draw[->] (b) edge (d);
  \draw[->] (b1) edge (f);
  \draw[->] (c) edge (e);
  \draw[->] (c) edge (f1);
  \draw[->] (d) edge (f2);
  \draw[->] (d) edge (g);
  \draw[->] (e) edge (h);
  \draw[->] (e) edge (i);
  \draw[->] (f) edge (j);
  \draw[->] (f) edge (k);
  \draw[->] (g) edge (l);
  \draw[->] (g) edge (m);
  \draw[->] (h) edge (n);
  \draw[->] (h) edge (o);
  \draw[->] (i) edge (p);
  \draw[->] (i) edge (q);
  \draw[->] (j) edge (r);
  \draw[->] (j) edge (s);
  \draw[->] (k) edge (t);
  \draw[->] (k) edge (u);
  \draw[->] (l) edge (v);
  \draw[->] (l) edge (w);
  \draw[->] (m) edge (x);
  \draw[->] (m) edge (y);
\end{tikzpicture}
\caption{Dividing Fig \ref{multisource} into two single source trees and a multi-source tree} \label{graphview}
\end{center}
\end{figure}
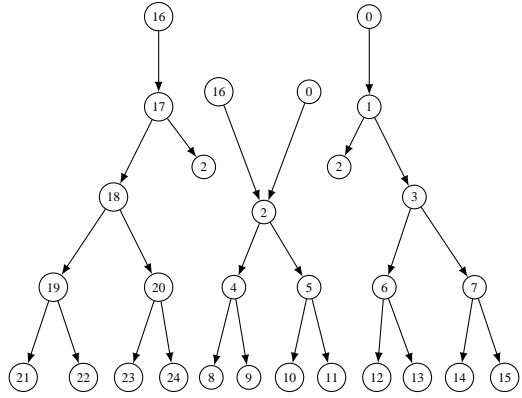

   A d-tree in the shape of tree is an intersection of the multicast trees that carry probes from sources to receivers. To retain the dispatching scenario, a virtual link is added to connect  a source sending probes to the intersection that results in  a new structure called multi-source tree. For instance, the general network shown in Fig. \ref{multisource} can be divided into three d-trees. Adding two virtual links to connect the two sources to the intersection rooted at node 2, we have a multi-source tree and two multicast trees as Fig. \ref{graphview}.  It is clear that the estimates of the two multicast trees in Fig. \ref{graphview} depend on the estimate of the multi-source tree. Thus, we need to estimate the pass rate of the multi-source tree first, and then use the pass rate to estimate the pass rates of the links on the upstream of the multi-source tree, such as the link connecting node $17$ to node $2$ in Fig. \ref{graphview}. The estimation order ensures the estimates obtained for the upstream links are   consistent with that in the downstream.

\subsection{Statistics and Likelihood Function of Multi-source Tree}

To have  a likelihood function describing the passing process from source $s$ to $R(i)$, we need a statistic that is defined as the number of probes sent by source $s$ and observed by $R(i)$, i.e.

\begin{equation}
n_i(s)=\sum_{o=1}^{n^s} Y_i^s(o), s \in S, i \in V.
\label{nk1}
\end{equation}

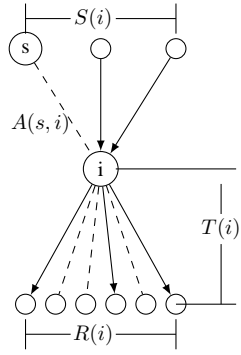
\begin{figure}
\begin{center}
\begin{tikzpicture}[scale=0.2,every path/.style={>=latex},every node/.style={scale=0.8}]
 \node[draw,circle]            (a) at (2,25)  { s };
 \node[draw,circle]            (l) at (7,25)  {  };
   \node[draw,circle]            (b) at (12,25)  {  };
  \node[draw,circle]            (c) at (7,17)  { i };
  \node[draw,circle]            (e) at (2,8)  {  };
  \node[draw,circle]            (f) at (4,8) {  };
  \node[draw,circle]            (g) at (6,8)  {  };
  \node[draw,circle]            (h) at (8,8) {  };
  \node[draw,circle]            (i) at (10,8) {  };
 \node[draw,circle]            (j) at (12,8)  {  };
  \draw (8,17) -- (16,17);
  \draw[dashed] (a) edge (c);
  \draw[->] (b) edge (c);
  \draw[->] (l) edge (c);
  \draw[->] (c) edge (e);
  \draw[dashed] (c) edge (f);
  \draw[dashed] (c) edge (g);
  \draw[->] (c) edge (h);
  \draw[dashed] (c) edge (i);
  \draw[->] (c) edge (j);
 \draw (12,8) -- (16,8);
 \draw (15,8) -- (15,12);
 \draw (15,14) -- (15,16);
 \draw (2,27) --(5,27);
 \draw (8,27) --(12,27);
 \draw (2,28) --(2,26);
 \draw (12,28) --(12,26);
\node at (6.5,27) {$S(i)$};
 \draw (2,6) --(5,6);
 \draw (8,6) --(12,6);
 \draw (2,7) --(2,5);
 \draw (12,7) --(12,5);
\node at (6.5,6) {$R(i)$};
 \node at (15,13) {$T(i)$};
 \node at (3,20) {$A(s,i)$};
\end{tikzpicture}
\caption{Multi-source tree} \label{multisource tree}
\end{center}
\end{figure}

Based on the assumption made at the beginning of this section, the probing process for intersection $i$, i.e. from $S(i)$ to $R(i)$, can be modelled as a Bernoulli process and illustrated as Fig. \ref{multisource tree}, where  $s,\mbox{ } s\in S(i)$ sends probes to $R(i)$ via node $i$ independently.  Apart from the symbols defined above, a few more are introduced here for the likelihood function of the multi-source tree created for intersection $i$, where $\Omega(i)=\{\Omega_s(i)|s \in S(i)\}$ is used to denote the observation of $R(i)$,  $A(s,i)$ is used to denote the pass rate of the path connecting $s, s \in S(i)$ to node $i$, and $\beta_i$ denotes the pass rate of $T(i)$. To ease the following discussion, we call the multi-source tree built on intersection $i$ multi-source tree $i$. We then have the likelihood function for multi-source tree $i$ as follows:
\begin{eqnarray}
{\cal{L}}(\beta_i, A(s,i)|\Omega(i))=\prod_{s\in S(i)} \Big
[(A(s,i)\beta_i)^{n_i(s)} \times  (1-A(s,i)\beta_i)^{(n^s-n_i(s))}\Big]
\label{likelihood equation}
\end{eqnarray}
 where $\beta_i$ is considered to be the unknown parameter generating $\Omega(i)$ since $A(s,i)$ is related to $\beta_i$ by $A(s, i)\beta_i = \gamma_i(s)$. $\gamma_i(s)$ is the pass rate from source $s$, via node $i$,  to $R(i)$ that can be estimated from observation, i.e. $\hat\gamma_i(s) =\frac{n_i(s)}{n^s}$. Obviously, (\ref{likelihood equation})  considers all of the probes sent by $S(i)$.

\subsection{Sufficient Statistics} \label{proof sufficient}

In contrast to the tree topology, the sufficient statistic of the likelihood function as (\ref{likelihood equation}) consists of a vector of $n_i(s), s \in S(i), i \in J$.  Let $t_i$ denote the sufficient statistic, we have
\begin{eqnarray}
t_i&=&\{n_i(s) | s \in S(i)\}
\label{intersectionnk1}
\end{eqnarray}
Rather than using the factorisation theorem, we use the mathematic definition of a sufficient statistic
to prove the sufficiency of $t_i$ in the following theorem.

\begin{theorem}\label{complete minimal sufficient statistics}
Let $Y_i^s=\{Y_i^s(1),....,Y_i^s(n^s)\}$ be the random sample obtained by $R(i)$ for the
 probes sent by $s$ and let $p_{\beta_i}(Y_i^s)$ be  the probability function of the passing process that generates $Y_i^s$.  Similarly,
 let $\prod_{s \in S(i)}p_{\beta_i}(Y_i^s)$ be the probability function of the passing process of $Y_i=\{Y_i^s | s \in S(i)\}$ from $S(i)$ to $R(i)$.  $t_i$ is minimal
sufficient statistic for $\beta_i$.
\end{theorem}

\begin{IEEEproof}
According to the definition of sufficiency, we need to prove
\begin{equation}\label{suffcondition}
\prod_{s \in S(i)}p_{\beta_i}(Y_i^s | n_i(s))=
\prod_{s \in S(i)}\frac{p_{\beta_i}(Y_i^s)}{p_{\beta_i}(n_i(s),Y_i^s)}
\end{equation}
is independent of $\beta_i$, that  can be achieved in three steps.

\begin{enumerate}
\item From (\ref{likelihood equation}), we have:

\begin{equation}\label{likelihood0}
p_{\beta_i}(Y_i^s)=(A(s,i)\beta_i)^{n_i(s)}
(1-A(s,i)\beta_i)^{n^s-n_i(s)}.
\end{equation}

\item Considering $Y_i^s$ the sample space that has $n_i(s) $, we have a binomial distribution
\begin{equation}
p_{\beta_i}(n_i(s), Y_i^s)= \\ \binom{n^s}{n_i(s)}(A(s,i)\beta_i)^{n_i(s)}(1-A(s,i)\beta_i)^{n^s-n_i(s)}.
\end{equation}
and
\begin{equation}
p_{\beta_i}(Y_i^s|n_i(s))= \\ \dfrac{(A(s,i)\beta_i)^{n_i(s)}(1-A(s,i)\beta_i)^{n^s-n_i(s)}}{\binom{n^s}{n_i(s)}(A(s,i)\beta_i)^{n_i(s)}(1-A(s,i)\beta_i)^{n^s-n_i(s)}
} \\
 =\dfrac{1}{\binom{n^s}{n_i(s)}}.
\end{equation}
\item Given the above, we have
\begin{equation}
\prod_{s \in S(i)} p_{\beta_i}(Y_i^s|n_i(s) )
=\prod_{s \in S(i)}\dfrac{1}{\binom{n^s}{n_i(s)}},
\end{equation}
\end{enumerate}
which is independent of $\beta_i$. Then, $t_i$ is the sufficient statistic to the multi-source tree.

Apart from the sufficiency,
 $n_i(s)$,  as defined in (\ref{nk1}), is the number of the probes, sent by $s$,  reaching $R(i)$ from $Y_i^s$. Since each probe is counted once and once only regardless of how many receivers in $R(i)$ observe the probe,  $t_i$ is a minimal sufficient statistic.
\end{IEEEproof}

\section{General Topology and its Estimator}
\label{section4}

Before deriving an estimator for multi-source tree $i$,  a lemma is needed that expresses $A(s,i), s \in S(i)$ by $\beta_i$ in order to have  a likelihood equation with single parameter rather than a equation with multiple parameters.
In addition, rather than using
  $A(s,i), s \in S(i)$ as the parameter to be estimated as that in the tree topology, we consider $\beta_i$ the parameter  to be estimated and have the following lemma to connect $A(s, i), s \in S(i)$ to $\widehat\beta_i$.

\begin{lemma} \label{beta MLE}
The maximum likelihood estimate of $\beta_i, i \in J$ is equal to
\begin{eqnarray}
\widehat\beta_i=\dfrac{\sum_{s \in S(i)} n_i(s)}{n_i^*}
\label{intersectionestimate}
\end{eqnarray}
where
\[
n_i^*=\sum_{s \in S(i)}n^{s}A(s, i),
\]
is the number of probes reaching node $i$.
\end{lemma}
\begin{IEEEproof}
Given the likelihood function presented in (\ref{likelihood equation}), it is easy to convert it into the log-likelihood function as follows:

\begin{eqnarray}
L(\beta_i,A(s,i)|\Omega(i))=\sum_{s \in S(i)} \Big[ n_i(s)(\log A(s,i) +\log \beta_i)+(n^s-n_i(s))\log(1-A(s,i)\beta_i)\Big ].
\label{pathlikely}
\end{eqnarray}

\noindent Differentiating the log-likelihood function with respect to ({\it wrt})
$\beta_i$ and letting the derivative be 0, we have
\begin{eqnarray}
\dfrac{\partial (L(\beta_i,A(s,i)|\Omega(i)))}{\partial \beta_i}&=&  \sum_{s \in S(i)}\Big(n_i(s)\beta_i^{-1}-(n^s-n_i(s))A(s,i)(1-A(s,i)\beta_i)^{-1}\Big) \nonumber \\
&=&\sum_{s \in S(i)}\Big(n_i(s)\beta_i^{-1}-(n^s-n_i(s)) A(s,i)(1-\gamma_i(s))^{-1}\Big) \nonumber\\
&=&\sum_{s \in S(i)}\Big(n_i(s)\beta_i^{-1}-(n^sA(s,i)\Big) =0. \mbox{ \vspace{2cm}     }
\label{derivative}
\end{eqnarray}
(\ref{derivative}) leads to
(\ref{intersectionestimate}). The passing process described by (\ref{likelihood equation}) follows Bernoulli distribution, a member of the exponential family with $\beta_i$
as the natural parameters. Then, based on the results about the exponential
families, such as \cite{Brown86}, $\hat\beta_i$ is the MLE
of $\beta_i$.
\end{IEEEproof}

\subsection{Estimator for Multi-source Tree}

Given lemma  \ref{beta MLE}, we derive a MLE for $\beta_i$ that considers all available information in the following theorem.

\begin{theorem}\label{general MLE}
Let $\beta_i$ be the pass rate of intersection $i, i \in J$. There is a polynomial,
$H(\beta_i, S(i))$, as follows:
\begin{enumerate}
\item \begin{eqnarray}
H(\beta_i, S(i)) &&= 1-\beta_i - \prod_{j\in d_i}
  \Big(1-\dfrac{\sum_{s' \in S(i)} n_j(s')}{\sum_{s \in S(i)}n_i(s)}\beta_i\Big) =0. \nonumber \\ \label{mgeneralpolybeta}
\end{eqnarray}
\item There is a unique solution to the polynomial if
\[
\prod_{j \in d_i}  \Big(1-\dfrac{\sum_{s' \in S(i)} n_j(s')}{\sum_{s \in S(i)}n_i(s)}\beta_i\Big) >1.
\]
\end{enumerate}
\end{theorem}
\begin{IEEEproof}
 Let $\alpha_j$ be the pass rate of the link that ends at node $j$. We then have:
\begin{eqnarray}
1-\beta_i&=&\prod_{j \in d_i} (1-\alpha_j\beta_j).
\label{betai expression}
\end{eqnarray}
where the left hand side is the loss rate of $T(i)$ and the right hand side (RHS) is the product of $TL(j), j \in d_i$.

According to lemma \ref{beta MLE}, we have
\begin{equation}
\dfrac{\hat\alpha_j\widehat\beta_j}{\widehat\beta_i}=\dfrac{\dfrac{\sum_{s \in S(i)} n_j(s)}{n_i^*}}{\dfrac{\sum_{s \in S(i)} n_i(s)}{n_i^*}}=\dfrac{\sum_{s \in S(i)} n_j(s)}{\sum_{s \in S(i)} n_i(s)}.
\label{sub2parent}
\end{equation}
Putting (\ref{sub2parent}) into (\ref{betai expression}), we have
\begin{eqnarray}
H(\beta_i, S(i)) &=& 1-\beta_i - \prod_{j\in d_i}
  \Big(1-\dfrac{\sum_{s' \in S(i)} n_j(s')}{\sum_{s \in S(i)}n_i(s)}\beta_i\Big) \nonumber \\&=&0.
  \label{unique solution}
\end{eqnarray}
 (\ref{unique solution}) is a function of $\beta_i$ in $[0,1]$. Since $H(0, S(i))=H(1,S(i))$,  (\ref{unique solution}) is solvable according to the Rolle's theorem. Then, we have the second point.
\end{IEEEproof}
Thus, $\hat A(s,i)$ can be obtained by $\frac{\gamma_i(s)}{\hat \beta_i}$, where $\hat \beta_i$ is obtained from (\ref{mgeneralpolybeta}).
Let $n_i=\sum_{s \in S(i)}n_i(s)$,  (\ref{mgeneralpolybeta}) can be written as
\begin{equation}
H(\beta_i, S(i))=1-\beta_i-\sum_{j \in d_i}(1-\frac{n_j}{n_i}\beta_i).
\end{equation}
 This is equivalent to transfer a multi-source tree to a multicast tree as illustrated in Fig. \ref{Transformation}, where all of $s, s \in S(i),$ are summarized into a virtual source (VS) that uses a virtual path to connects node $i$.

\subsection{Estimator of Upstream Link}
\label{4-2}

 As stated, the $\hat\beta_i$ obtained by (\ref{mgeneralpolybeta}) for multi-source tree $i$ can be different from that obtained by an estimator developed for the tree topology. To be consistent between the estimates obtained for the d-trees that neighbour each other, an estimator differing from that developed for the tree topology is needed that has $\hat\beta_i$ embedded in it.
 Such an estimator can be derived from (\ref{mgeneralpolybeta}) as:
 \begin{eqnarray}
H(A(s,f^s(i)), s)=1-\frac{\gamma_{f^s(i)}(s)}{A(s, f^s(i))}-\prod_{k \in d_{f^s(i)}}(1-\frac{\gamma_k(s)}{A(s,f^s(i))} ).
\label{tree likelihood equation}
\end{eqnarray}
 To take into account the impact of the $\hat\beta_i$ obtained from (\ref{mgeneralpolybeta}) on the estimate of $A(s,f^s(i))$, $\hat A(s,i)$ is used to replace $\gamma_i(s)$, we then have:
\begin{eqnarray}
H(A(s,f^s(i)), s)=1-\dfrac{\gamma_{f^s(i)}(s)}{A(s, f^s(i))}-\big[1-\dfrac{\hat A(s,
i)}{A(s, f^s(i))}\big] \prod_{k \in d_{f^s(i)} \setminus
i}(1-\dfrac{\gamma_k(s)}{A(s,f^s(i))} ).
 \label{rateestimator}
\end{eqnarray}
Except $A(s, f^s(i))$ that is the parameter to be estimated, all others in (\ref{rateestimator}) are either obtained by $\dfrac{\hat\gamma_i(s)}{\hat\beta_i}$ $(\hat A(s,i))$ or replaced by the corresponding empirical values. (\ref{rateestimator}) is the MLE given the pass rate of the path from $s$ to node $i$ is $\hat A(s,
i)$.
Once
having $\hat A(s, f^s(i))$, we can move a level up toward $s$ and use (\ref{rateestimator}) again to
estimate $A(s, f^s_2(i))$. This process continues until
reaching $s$. If
 the tree being estimated has more than one intersections, at
the common ancestor of the intersections, the RHS of
(\ref{rateestimator}) will have a number of the left-most terms, one
for an intersection plus the product term for those that only receive probes from $s$. If the total number of intersections plus
the number of independent subtrees is greater than 5, there is no
closed form solution to (\ref{rateestimator}). Using the Rolle's theorem, we can prove there is only one solution in $(0,1]$ in (\ref{rateestimator}) that is the maximum likelihood estimate. Therefore, (\ref{mgeneralpolybeta}) and  (\ref{rateestimator}), plus the estimation order, consist of  the maximum likelihood estimator of the general topology that uses all available information and maintains the consistence between estimates.

%

\section{Statistical Property of the Estimators}
\label{section5}

  (\ref{mgeneralpolybeta}), as the likelihood equation of multi-source tree $i$,  sets up the foundation of loss tomography in the general topology. To understand (\ref{mgeneralpolybeta}), a statistical analysis is carried out to identify the statistical properties of the estimator. In addition, we have the statistical properties of the MVWA. This section is devoted to the properties.

\subsection{Generalization}
\label{comparison}
\begin{theorem}\label{special case}
 (\ref{mgeneralpolybeta}) is applicable to  the tree topology.
\end{theorem}
\begin{IEEEproof} Let $\cal{N}=\{T\}$ and there is a source $s$ attached to the root of $\cal T$. For  node $i, i \notin R$, we have $S(i)=\{s\}$. Then, (\ref{mgeneralpolybeta}) is transformed to:
\begin{eqnarray}
H(\beta_i, S(i)) &=& 1-\beta_i -
\prod_{j \in d_i}(1-\dfrac{\sum_{k \in S(i)} n_j(k)}{ \sum_{k \in S(i)} n_i(k)}\beta_i) \nonumber \\
&=&1-\beta_i -\prod_{j \in
d_i}(1-\dfrac{ n_j(s)}{n_i(s)}\beta_i) \label{single source} \\
&=&1-\dfrac{\gamma_i(s)}{A(s,i)} -\prod_{j \in
d_i}(1-\dfrac{\gamma_j(s)}{A(s,i)})=0. \label{tree MLE}
\end{eqnarray}

\noindent If the $s$ occurred in the last equation refers to the single source of the tree topology, we have an equation that is equivalent to  $H(A_i, i)$ presented in
\cite{CDHT99}.
\end{IEEEproof}

%
%
%

\subsection{ Minimum-Variance Unbiased Estimator (MVUE)}
It is known that if a MLE is a function of the sufficient statistic, it is asymptotically unbiased, consistent and asymptotically efficient. Thus,  the estimate obtained by (\ref{mgeneralpolybeta}) has all of these properties.
In addition, it is a MVUE.
\begin{theorem} \label{MVUE theorem}
The estimator proposed for intersections is a MVUE and the variances of the
estimates reach the Cram\'{e}r-Rao bound.
\end{theorem}
\begin{IEEEproof}
The proof is based on Rao-Blackwell Theorem \cite{UC08} that states if $g(X)$
is any kind of estimator of a parameter $\theta$, then the conditional
expectation of $g(X)$ given $T(X)$, where $T(X)$ is a sufficient statistic, is
typically a better estimator of $\theta$, and is never worse. Further,
if the estimator is the only unbiased estimator based on $T(X)$, then, the estimator
is the MVUE.

We can prove $\widehat\beta_i, i  \in J$ obtained by (\ref{mgeneralpolybeta}) is an unbiased estimate from (\ref{intersectionestimate}), where
the denominator is the total number of probes reaching node $i$ and the nominator is the total number of probes observed by $R(i)$. Clearly, $\widehat \beta_i$ is the sample mean that can be written as
\begin{equation}
{\widehat\beta_i} = \dfrac{\sum_{s \in S(i)} n_i(s)}{n_{i}^*}
\end{equation}
where $n_{i}^*=\sum_{s \in S(i)} n^s A(s,i)$ is the number of probes reaching node $i$.  We then have
\begin{eqnarray}
&&E\Big ( \dfrac{\sum_{s \in S(i)} n_i(s)}{n_{i}^*}\Big ) \nonumber \\
&=& E\Big( \dfrac{1}{n_{i}^*}\sum_{j=1}^{n_{i}^*}x_j\Big) \nonumber \\
&=& \beta_i
\end{eqnarray}
Considering the assumption made on the loss process of a link or a path, i.e. it follows Bernoulli distribution, $\widehat \beta_i$ is a unbiased estimate of $\beta_i$. In addition, the statistics used in estimation has been proved to be the sufficient statistics.  Applying Rao-Blackwell theorem, we have
the first part of the theorem.
Based on the results presented in \cite{Joshi76}, it is easy to prove the variance of
the estimates are equal to Cram\'{e}r-Rao low bound since
(\ref{pathlikely}), the likelihood function, falls into the standard
exponential family.
\end{IEEEproof}

 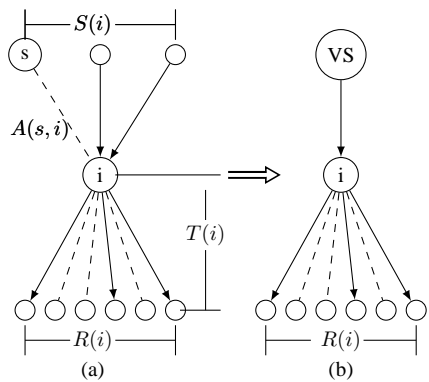
\begin{figure}
\begin{center}
\begin{tikzpicture}[scale=0.2,every path/.style={>=latex},every node/.style={scale=0.8}]
 \node[draw,circle]            (a) at (2,25)  { s };
 \node[draw,circle]            (l) at (7,25)  {  };
   \node[draw,circle]            (b) at (12,25)  {  };
  \node[draw,circle]            (c) at (7,17)  { i };
  \node[draw,circle]            (e) at (2,8)  {  };
  \node[draw,circle]            (f) at (4,8) {  };
  \node[draw,circle]            (g) at (6,8)  {  };
  \node[draw,circle]            (h) at (8,8) {  };
  \node[draw,circle]            (i) at (10,8) {  };
 \node[draw,circle]            (j) at (12,8)  {  };
  \draw (8,17) -- (15,17);
  \draw[dashed] (a) edge (c);
  \draw[->] (b) edge (c);
  \draw[->] (l) edge (c);
  \draw[->] (c) edge (e);
  \draw[dashed] (c) edge (f);
  \draw[dashed] (c) edge (g);
  \draw[->] (c) edge (h);
  \draw[dashed] (c) edge (i);
  \draw[->] (c) edge (j);
 \draw (12.3,8) -- (15,8);
 \draw (14,8) -- (14,12);
 \draw (14,14) -- (14,16);
 \draw (2,27) --(5,27);
 \draw (8,27) --(12,27);
 \draw (2,28) --(2,26);
 \draw (12,28) --(12,26);
\node at (6.5,27) {$S(i)$};
 \draw (2,6) --(5,6);
 \draw (8,6) --(12,6);
 \draw (2,7) --(2,5);
 \draw (12,7) --(12,5);
\node at (6.5,6) {$R(i)$};
\node at (6.5,4) {(a)};
 \node at (14,13) {$T(i)$};
 \node at (3,20) {$A(s,i)$};
 \node[draw,circle]            (l) at (23,25)  { VS};
  \node[draw,circle]            (c) at (23,17)  { i };
  \node[draw,circle]            (e) at (18,8)  {  };
  \node[draw,circle]            (f) at (20,8) {  };
  \node[draw,circle]            (g) at (22,8)  {  };
  \node[draw,circle]            (h) at (24,8) {  };
  \node[draw,circle]            (i) at (26,8) {  };
 \node[draw,circle]            (j) at (28,8)  {  };
  \draw[->] (l) edge (c);
  \draw[->] (c) edge (e);
  \draw[dashed] (c) edge (f);
  \draw[dashed] (c) edge (g);
  \draw[->] (c) edge (h);
  \draw[dashed] (c) edge (i);
  \draw[->] (c) edge (j);
 \draw (2,27) --(5,27);
 \draw (8,27) --(12,27);
 \draw (2,28) --(2,26);
 \draw (12,28) --(12,26);
\node at (6.5,27) {$S(i)$};
 \draw (18,6) --(21,6);
 \draw (24.5,6) --(28,6);
 \draw (18,7) --(18,5);
 \draw (28,7) --(28,5);
\node at (23,6) {$R(i)$};
\node at (23, 4) {(b)};
 \node at (3,20) {$A(s,i)$};
   \draw[vecArrow](15.5,17) to (19,17);
\end{tikzpicture}
\caption{Transformation} \label{Transformation}
\end{center}
\end{figure}

\subsection{Efficiency}

 Apart from knowing  the estimates obtained by (\ref{mgeneralpolybeta}) reaches the Cram\'{e}r-Rao low bound, we would like to know the low bound and are interested in the difference between the low bound and that obtained from direct measurement. Instead of deriving the low bound directly from the estimator, we want to achieve this via efficiency and have the following theorem for the Fisher information of (\ref{mgeneralpolybeta}).

\begin{theorem} \label{fisher info}
The Fisher information of a probe $y$ sent by $s$ on $\beta_i$ is equal to
\begin{equation}
\mathcal{I}(\beta_i|s, y)=\dfrac{A(s,i)}{\beta_i(1-A(s,i)\beta_i)}
\label{fisher s}
\end{equation}
\end{theorem}
\begin{IEEEproof}
The likelihood function of $y$ is equal to

\begin{eqnarray}
L(\beta_i,A(s,i)|y)&=& y(\log A(s,i) +\log \beta_i) +(1-y)\log(1-A(s,i)\beta_i). \nonumber
\label{fisher likelihood}
\end{eqnarray}
In addition,
\begin{eqnarray}
\dfrac{\partial^2 L(\beta_i,A(s,i)|y)}{\partial \beta_i^2}=-\dfrac{y}{\beta_i^2}-\dfrac{(1-y)A(s,i)^2}{(1-A(s,i)\beta_i)^2}. \nonumber
\end{eqnarray}
Then, we have
\begin{eqnarray}
E(-\dfrac{\partial^2 L(\beta_i,A(s,i)|y)}{\partial \beta_i^2})&=&\dfrac{E(y)}{\beta_i^2}+\dfrac{(E(1-y))A(s,i)^2}{(1-A(s,i)\beta_i)^2} \nonumber \\
&=&\dfrac{A(s,i)}{\beta_i(1-A(s,i)\beta_i)}. \nonumber
\end{eqnarray}
The theorem follows.
\end{IEEEproof}
(\ref{fisher s}) shows the information provided by a probe sent by source $s$ is proportional to the pass rate of the path connecting source $s$ to node $i$. If $n^s$ probes are sent by source $s$, the Fisher information is equal to
 \[
 \frac{ n^s A(s,i)}{\beta_i(1-A(s,i)\beta_i)}.
 \]

 If $A(s, i)=1$, $\mathcal{I}(\beta_i|s, y)=\dfrac{1}{\beta_i(1-\beta_i)}$ that equals to the information provided by direct measurement. In addition, given $\beta_i$ the efficiency is a monotonic function of $A(s, i)$ in $(0, 1]$.


\subsection{Variance}
Given (\ref{fisher s}), we have the following theorem for the variance of $\hat\beta_i$ in regard to the probes sent by $s$.
\begin{theorem} \label{variance}
Let $var_s(\hat\beta_i|y)$ be the variance of the estimates obtained by (\ref{mgeneralpolybeta}) in regards to $y$ sent by $s, s \in S(i)$ that is equal to
\begin{equation}
var_s(\hat\beta_i|y)=\dfrac{\beta_i(1-A(s,i)\beta_i)}{A(s,i)}
\label{variance beta}
\end{equation}
\end{theorem}
\begin{IEEEproof}
 As stated, the passing process is a Bernoulli process that falls into the exponential family and the likelihood function satisfies the regularity conditions presented in \cite{Joshi76}. Thus, the variance of  the estimator reaches the Cram\'{e}r-Rao bound that is the reciprocal of the Fisher information.
\end{IEEEproof}

Theorem \ref{fisher info} and \ref{variance} show the impact of $A(s,i)$ on the efficiency of (\ref{mgeneralpolybeta}) and the variance of the estimates obtained by (\ref{mgeneralpolybeta}), respectively. Further, we have the corollary for the Fisher information and the variance of $\hat\beta_i$ obtained by (\ref{mgeneralpolybeta}) in  an experiment.
\begin{corollary} \label{Fisher corollary}
The Fisher information obtained from $Y$ for $\beta_i$ is equal to
\begin{equation}
\mathcal{I}(\beta_i|S(i), Y)= \sum_{s \in S(i)}\dfrac{n^s A(s,i)}{\beta_i(1-A(s,i)\beta_i)}
\label{information Y}
\end{equation}
and the variance of the estimate obtained from $Y$
\begin{equation}
var(\hat\beta_i|Y)= \Big (\sum_{s \in S(i)}\dfrac{n^s A(s,i)}{\beta_i(1-A(s,i)\beta_i)}\Big)^{-1}
\label{variance Y}
\end{equation}
\end{corollary}
\begin{IEEEproof}
Summing the information for each probe we have (\ref{information Y}) because the Fisher information is additive and the {\it i.i.d.} assumption made at \ref{section3.a},  . In addition, because the variance of the sum is the sum of the variances, we have (\ref{variance Y}).
\end{IEEEproof}

\subsection{Low Bound of MVWA and Alternatives}
\label{principle}
As shown in Fig. \ref{Transformation}, (\ref{mgeneralpolybeta}) differs from (\ref{tree MLE}) by switching the parameter to be estimated from $A(s,i)$ to $\beta_i$ that makes (\ref{mgeneralpolybeta}) capable of considering all of the probes sent by $S(i)$ to $T(i)$.
Because of the switch, the estimate obtained by (\ref{mgeneralpolybeta}), in terms of efficiency, is better than that obtained by other estimators. Apart from that, the variance achieved by (\ref{mgeneralpolybeta}) is the low bound of the MVWA. Let $var_{MA}(\cdot)$ be the variance of the MVWA, we have a corollary to prove (\ref{variance Y}) is the low bound of the MVWA.
\begin{corollary} \label{MVWA corollary}
The low bound of the variance of the estimate obtained by the MVWA is equal to

\begin{equation}
var_{MA}(\hat\beta_i|Y)= \Big (\sum_{s \in S(i)}\dfrac{n^s A(s,i)}{\beta_i(1-A(s,i)\beta_i)}\Big)^{-1}
\label{variance MVWA}
\end{equation}
\end{corollary}
\begin{IEEEproof}
The MVWA uses
\begin{equation}
\hat \beta_i =	\sum_{s \in S(i)} \lambda_s \hat\beta_i(s), 	\lambda_s \in [0,1];	\sum_{s \in S(i)}\lambda_s =1.
\label{MA condition}
\end{equation}
to estimate the pass rate of multi-source tree $i$,  where $\hat\beta_i(s)$ and $\lambda_s$ are the pass rate estimated from the probes sent by $s$ and the weight assigned to  $\hat\beta_i(s)$, respectively. Then,
\begin{equation}
var_{MA}(\hat\beta_i|Y)=\sum_{s \in S(i)} \lambda_s^2 var_s(\hat\beta_i|Y_s)
\label{MVWA variance}
\end{equation}
Given the condition specified in (\ref{MA condition}),  we use the method of Lagrange multiplier to find $\lambda_s$ that can minimize (\ref{MVWA variance}). The solution is:
 \begin{equation}
 \lambda_s=\frac{\frac{1}{var_s(\hat\beta_i|Y_s)}}{\sum_{s \in S(i)} \frac{1}{var_s(\hat\beta_i|Y_s)}}
 \label{lambda MVWA}
 \end{equation}
where
\begin{equation}
var_s(\hat\beta_i|Y_s)=\Big(\frac{n^s A(s,i)}{\beta_i(1-A(s,i)\beta_i)}\Big)^{-1}
\label{variance s}
\end{equation}
Using the above in (\ref{MVWA variance}), we have (\ref{variance MVWA}).
\end{IEEEproof}

  Although it is impossible to know $var_s(\hat\beta_i|Y_s)$ in advance, the derivation shows that the accuracy of the estimate obtained by the MVWA depends on the accuracy of $\lambda_s$ that further depends on the number of probes sent by $s$ since $var_s(\hat\beta_i|Y_s)=\frac{\beta_i(1-A(s,i)\beta_i)}{n^s A(s,i)}$. Note that the low bound can only be achieved if the variances are known in advance that is equivalent to have extra information. Thus, the MLE proposed in this paper is better than the MVWA.

 Using (\ref{lambda MVWA}) and (\ref{variance s}), we can improve the MVWA by removing the need of advanced knowledge about the variances and have an alternative. The following corollary shows the details of the alternative.

 \begin{corollary} \label{RBMVWA corollary}
A  renewed MVWA, called RBMVWA, has the same low bound for the variance of its estimates as that of the MLE without the need to know the variances in advance. The RBMVWA relies on end-to-end observation to compute the weight for each of the estimates obtained by  the MLE designed for the tree topology, where the weight is equal to
\begin{equation}
\lambda_s=\frac{\frac{\gamma_i(s)}{1-\gamma_i(s)}}{\sum_{s \in S(i)} \frac{\gamma_i(s)}{1-\gamma_i(s)}}, \mbox{  } s \in S(i).
\label{RBMVWA lambda}
\end{equation}
\end{corollary}
\begin{IEEEproof}
Using (\ref{variance beta}) in (\ref{lambda MVWA}), we have
\begin{eqnarray}
\lambda_s&=&\frac{(\frac{A(s,i)}{\beta_i(1-A(s,i)\beta_i)})}{\sum_{s \in S(i)} {(\frac{ A(s,i)} {\beta_i(1-A(s,i)\beta_i)})}}.
\end{eqnarray}
Multiplying $\beta_i^2$ to the nominator and denominator, we have
\begin{eqnarray}
\lambda_s&=&\frac{\frac{\gamma_i(s)}{(1-\gamma_i(s))}}{\sum_{s \in S(i)} {\frac{ \gamma_i(s)} {(1-\gamma_i(s))}}} \nonumber
\end{eqnarray}
since $A(s,i)\beta_i=\gamma_i(s)$.
\end{IEEEproof}

Although the hurdle of advanced knowledge is removed,  the RBMVWA still needs to send a large number of probes from $s, s \in S(i)$ to $R(i)$ to make $\frac{\gamma_i(s)}{(1-\gamma_i(s))}$ stable since the function increases exponentially if $\hat\gamma_i(s) \rightarrow 1$.  To reduce the sensitivity and increase the robustness of the RBMVWA, we have another alternative for the RBMVWA that sets
$\lambda_s=\frac{\gamma_i(s)}{(\sum_{s \in S(i)} \gamma_i(s))}$, called modified RBMVWA, (MRBMVWA). Compared with RBMVWA, MRBMVWA scarifies sensitivity for robustness. For instance, assume there are two sources, say $s$ and $t$, sending probes to multi-source tree $i$, if there is little difference between $\gamma_i(s)$ and $\gamma_i(t)$ the estimate obtained by MRBMVWA should be as good as that of the MLE; on the other hand, if there is noticeable difference between $\gamma_i(s)$ and $\gamma_i(t)$, the estimates may not match $\gamma_i(s)$ and $\gamma_i(t)$.

\subsection{Variance of Upstream Link}

In Section \ref{4-2}, (\ref{rateestimator}) is used to estimate the pass rate of a upstream path, where the estimate obtained by (\ref{rateestimator}) may be more accurate than that obtained by (\ref{tree MLE}) in theory since the $\hat \beta_i$ obtained by (\ref{mgeneralpolybeta}) has a smaller variance according to corollary \ref{Fisher corollary}. Using the same procedure as that used in corollary \ref{Fisher corollary}, we have
\begin{equation}
var(A(s,i)|y)=\frac{A(s,i)(1-A(s,i)\beta_i)}{\beta_i}.
\end{equation}
If $n^s$ probes are sent from $s$ to $R(i)$,
\begin{equation}
var(A(s,i)|Y)=\frac{A(s,i)(1-A(s,i)\beta_i)}{n^s\beta_i}.
\label{variance of A(s,i)}
\end{equation}
(\ref{variance of A(s,i)}) is identical with the variance obtained in \cite{Zhu15} for the path connecting the source to an internal node in the tree topology. Therefore, there is no significant advantage of (\ref{rateestimator}) over  (\ref{tree MLE}) unless the $\hat\beta_i$ obtained by (\ref{mgeneralpolybeta}) is very different from that obtained by (\ref{tree MLE}).

\section{Simulation}
\label{section8}

Given the statistical properties presented in the last section, we are able to evaluate some of the estimators proposed so far and reach some conclusions without running simulations or experiments. For instance, from Theorem \ref{variance}, we can conclude that the variance of $\hat\beta_i$ monotonically decreases as $A(s,i), s \in S(i)$ and the variance of $\hat A(s,i)$ monotonically decreases as $\beta_i$.
Because of this,
only a small set of simulations are conducted to verify the properties and test the effectiveness of the estimators discussed in this paper. The network used in the simulations is as Fig. \ref{multisource tree}, where two sources are used to multicast probes to a number of receivers connected to node $i$. The number of receivers attached to node $i$ varies from 2 to 3 and the loss rates of the links are either $1\%$ or $3\%$.

Given the above, a number of networks can be configured that consist of a network structure and a set of the loss rates assigned to each link. To represent the networks, we name each of them a setting that is specified by $n \mbox{ } \times \mbox{ }m$ and $(s1, \mbox{ } s2) \times (d1, \mbox{ } d2)$ or $(s1, \mbox{ } s2) \times (d1, \mbox{ } d2, \mbox{ } d3)$, where $n$ is for the number of sources, $m$ is for the number of receivers, $s1$ and $s2$ are the loss rates of the links connecting the two sources to node $i$, and $dj, j \in \{1,2,3\}$ is the loss rate assigned to the link connecting node $i$ to receiver $j$.   The number of probes sent by a source in a simulation varies in 100, 500, and 1000.

To have the mean and variance of a setting for a probe size, we run the same simulation 20 times for a setting and each time  a unique number is used as the seed of the random number generator. The statistics obtained by an estimator for a specific setting  is  presented in a table. Because of the difference between the estimator developed for the tree topology and the others, there are two type of tables. The table designated for the estimator developed for the tree topology has 12 columns, except the first column, as shown in Table \ref{tree2-2}.  In contrast to the table designated for the tree topology, the other table has 6 columns except the first one as Table \ref{proposed2-2}. Each of the tables, regardless of the type, has a number of rows. Apart from the first few rows that are used for headings, each row is for a link that is specified by the name in the first column, where {\it source 1} and {\it source 2} are for the links connecting the sources to node $i$ and {\it link  $j, \mbox{ }j \in \{1, 2, 3\}$} for the link connecting node $i$ to receiver $j$.

 The results for the setting $2 \times 2$ with $(0.01,0.01) \times (0.01,0.01)$ are presented in Table \ref{tree2-2} and Table \ref{proposed2-2}, where Table \ref{tree2-2} is for the estimator developed for the tree topology and Table \ref{proposed2-2} is for the proposed estimator. As expected,  the estimates obtained by the estimators become closer as the increase of the number of probes. In addition, there is slightly difference between the estimates obtained for the links connecting the sources to node $i$, where the loss rates of source 1 and source 2 are obtained in Table \ref{proposed2-2} that show the impact of $\hat\beta_i$ on the estimates of the upstream links if loss rate is lower and sample size is small. The differences in terms of the estimates are obvious for the links connecting node $i$ to receivers, in particular in variance, where the estimates obtained by the proposed estimator is much lower than that of the tree ones. Further, the variances of the estimates obtained by the estimator of the tree topology for the same link are different from each other until the number of probes sent reaches 1000. Because of this, it is risky to use the variances to determine the weights of the MVWA unless a large number of probes are sent. To compare RBMVWA with MRBMVWA  proposed in Section \ref{principle} we use the two estimators to estimate the loss rates of link 1 and link 2. The results are presented in Tables \ref{RBMVWA2-2} and \ref{modifiedRBMVWA2-2} that agree with the statistical properties identified in Section \ref{principle}.  Comparing the last two rows in Table \ref{RBMVWA2-2} with that in Table \ref{proposed2-2}, the difference is obvious if sample is less than 1000 that indicates the $\hat\gamma_i(s)$ obtained is not accurate enough as that obtained from sending 1000 probes. On the other hands, by comparing the last two rows of Table \ref{modifiedRBMVWA2-2} with that Table \ref{proposed2-2}, we find the results of the MRBMVWA are almost identical to that of the MLE since the pass rate of the path from a source to $R(i)$ is inversely proportional to the variance of the $\hat\beta_i$. The advantage of the MRBMVWA over the MVWA and the RBMVWA, as stated, is its robustness and is in favor of the situation that all sources in $S(i)$ has similar $\hat \gamma_i(s)$ since it uses a linear function to assign weights to estimates.

\begin{table*}[th]
  \centering
\scriptsize
\begin{tabular}{|c|r|r|r|r|r|r|r|r|r|r|r|r|} \hline
sample &\multicolumn{4}{|c|}{100 } & \multicolumn{4}{|c|}{500 } &\multicolumn{4}{|c|}{1000}	\\ \hline
Link & Mean & Var &	Mean & Var	& Mean & Var & Mean & Var &	Mean & Var	& Mean & Var\\ \hline
source 1 &	0.0069328 &	8.16E-05	&  &		&0.0096028	&1.72E-05	&  &		&0.0101894	&1.07E-05
&  & \\ \hline
source 2 &		&	&0.0088476	&6.94E-05	&  &		&0.0105106	&2.94E-05	&  &		&0.0104947	&1.06E-05
 \\ \hline
link 1	& 0.0111478	&1.54E-04	&0.0102473	&9.48E-05	&0.0104970	&1.42E-05	&0.0090832	&1.51E-05	&0.0107666	&1.07E-05	&0.0099073	&1.11E-05 \\ \hline
link 2	&0.0121581	&9.81E-05	&0.0137570	&8.53E-05	&0.0092935	&1.25E-05	&0.0100889	&1.29E-05	&0.0103663	&1.34E-05	&0.0108174	&1.34E-05\\ \hline
\end{tabular}
\caption{Estimate obtained by the tree estimator for 2 $\times$ 2 network, loss rates = $(1\%, 1\%) \times (1\%, 1\%)$ }
\label{tree2-2}
\end{table*}

\begin{table*}[th]
  \centering
\scriptsize
\begin{tabular}{|c|r|r|r|r|r|r|} \hline
sample &\multicolumn{2}{|c|}{100 } & \multicolumn{2}{|c|}{500 } &\multicolumn{2}{|c|}{1000}	\\ \hline
Link & Mean & Var &	Mean & Var	& Mean & Var\\ \hline
source 1 &	0.0068795 &	8.16E-05 &	0.0096058	&1.73E-05	&0.0101914	&1.07E-05 \\ \hline
source 2 &	0.0088800	&6.90E-05	&0.0105058	&2.94E-05	&0.0104914	&1.06E-05 \\ \hline
link 1	& 0.0107029	&7.32E-05	&0.0097883	&5.28E-06	&0.0103383	&4.89E-06 \\ \hline
link 2	&0.0129538	&2.92E-05	&0.0096917	&6.75E-06	&0.0105915	&2.38E-06 \\ \hline
\end{tabular}
\caption{Estimate obtained by the proposed estimators for 2 $\times$ 2 network, loss rates = $(1\%, 1\%) \times (1\%, 1\%)$}
\label{proposed2-2}
\end{table*}

\begin{table*}[th]
  \centering
\scriptsize
\begin{tabular}{|c|r|r|r|r|r|r|} \hline
sample &\multicolumn{2}{|c|}{100 } & \multicolumn{2}{|c|}{500 } &\multicolumn{2}{|c|}{1000}	\\ \hline
Link & Mean & Var &	Mean & Var	& Mean & Var\\ \hline
link 1 &0.00826228	&7.15E-05	&0.00921997	&5.11E-06	&0.0101077	&4.70E-06 \\ \hline
link 2 &0.00939153	&3.98E-05	&0.00936787	&7.79E-06	&0.0100406	&2.84E-06 \\ \hline
\end{tabular}
\caption{Estimate obtained by RAMVWA for 2 $\times$ 2 network, loss rates = $(1\%, 1\%) \times (1\%, 1\%)$}
\label{RBMVWA2-2}
\end{table*}

\begin{table*}[th]
  \centering
\scriptsize
\begin{tabular}{|c|r|r|r|r|r|r|} \hline
sample &\multicolumn{2}{|c|}{100 } & \multicolumn{2}{|c|}{500 } &\multicolumn{2}{|c|}{1000}	\\ \hline
Link & Mean & Var &	Mean & Var	& Mean & Var\\ \hline
link 1	& 0.0106505	&7.24E-05	&0.00977831	&5.27E-06	&0.0103321	&4.89E-06 \\ \hline
link 2	& 0.0128896	&2.89E-05	&0.00968556	&6.75E-06	&0.0105803	&2.37E-06 \\ \hline
\end{tabular}
\caption{Estimate obtained by MRAMVWA for 2 $\times$ 2 network, loss rates = $(1\%, 1\%) \times (1\%, 1\%)$}
\label{modifiedRBMVWA2-2}
\end{table*}

To investigate the impact of loss rates on the estimates, we vary the setting  by $(1\%, 3\%) \times (1\%, 3\%)$. The results are presented in Table \ref{tree2-2unequal} and Table \ref{proposed2-2unequal} that are almost identical as the previous ones except the variance is higher for the links that have higher loss rate. In addition, as expected the estimates obtained by the estimator developed for the tree topology are noticeable different from that obtained by the proposed estimator for all of links if 100 probes are sent by each source.

\begin{table*}[th]
  \centering
\scriptsize
\begin{tabular}{|c|r|r|r|r|r|r|r|r|r|r|r|r|} \hline
sample &\multicolumn{4}{|c|}{100 } & \multicolumn{4}{|c|}{500 } &\multicolumn{4}{|c|}{1000}	\\ \hline
Link & Mean & Var &	Mean & Var	& Mean & Var & Mean & Var &	Mean & Var	& Mean & Var\\ \hline
source 1 &0.00682135	&8.29E-05 & & 		&0.0095012	&1.75E-05	& &		&0.0102454	&1.06E-05	& &	
\\ \hline
source 2 &		&	&0.0351666	&1.62E-04	& &		&0.0333467	&7.49E-05	& &		&0.0309292	&3.08E-05
\\ \hline
link 1	&0.0112576	&1.55E-04 &0.0097113	&8.17E-05	&0.0105982	&1.45E-05	&0.0085287	&1.67E-05	&0.0107100	&1.22E-05	&0.0097718	&1.24E-05
 \\ \hline
link 2	&0.0298412	&4.60E-04 &0.0345327	&4.60E-04	&0.0285759	&4.19E-05	&0.030048	&5.67E-05	&0.0287000	&3.53E-05	&0.0291677	&3.73E-05
\\ \hline
\end{tabular}
\caption{Estimate obtained by the tree estimator for 2 $\times$ 2 network, loss rates = $(1\%, 3\%) \times (1\%, 3\%)$ }
\label{tree2-2unequal}
\end{table*}

\begin{table*}[th]
  \centering
\scriptsize
\begin{tabular}{|c|r|r|r|r|r|r|} \hline
sample &\multicolumn{2}{|c|}{100 } & \multicolumn{2}{|c|}{500 } &\multicolumn{2}{|c|}{1000}	\\ \hline
Link & Mean & Var &	Mean & Var	& Mean & Var\\ \hline
source 1 &	0.0066940	&8.24E-05	&0.0095170	&1.73E-05	&0.0102572	&1.04E-05
\\ \hline
source 2 & 0.0352047	&1.62E-04	&0.0333236	&7.45E-05	&0.0309134	&3.05E-05
 \\ \hline
link 1	&0.0104870	&6.81E-05	&0.0095826	&5.24E-06	&0.0102494	&5.32E-06
 \\ \hline
link 2	&0.0321491	&1.70E-04	&0.0292985	&2.67E-05	&0.0289328	&1.78E-05
 \\ \hline
\end{tabular}
\caption{Estimate obtained by the proposed estimators for 2 $\times$ 2 network, loss rates = $(1\%, 3\%) \times (1\%, 3\%)$ }
\label{proposed2-2unequal}
\end{table*}

To investigate the impact of the subtree connected at node $i$ on the estimates, we change the network configuration to $ 2 \times 3$ and the loss rates of the five links are set to $(1\%, 1\%) \times (1\%, 1\%, 1\%)$. The results are presented in  Table \ref{tree2-3} and  Table \ref{proposed2-3}. Comparing Table \ref{tree2-3} and  Table \ref{proposed2-3} with Table \ref{tree2-2} and  Table \ref{proposed2-2}, we observe the similarity and differences. The difference is the estimates obtained for source 1 and source 2 in Table \ref{tree2-3} and  Table \ref{proposed2-3} are more accurate than that in Table \ref{tree2-2} and  Table \ref{proposed2-2}. This agrees with (\ref{variance of A(s,i)}) that confirms the claim made at the beginning of this section.

\begin{table*}[th]
  \centering
\scriptsize
\begin{tabular}{|c|r|r|r|r|r|r|r|r|r|r|r|r|} \hline
sample &\multicolumn{4}{|c|}{100 } & \multicolumn{4}{|c|}{500 } &\multicolumn{4}{|c|}{1000}	\\ \hline
Link & Mean & Var &	Mean & Var	& Mean & Var & Mean & Var &	Mean & Var	& Mean & Var\\ \hline
source 1 &0.0089992	&4.90E-05	& &		&0.0106991	&1.77E-05 &		&	&0.010349	&1.50E-05	& &	
\\ \hline
source 2 &		&	&0.0094980	&7.47E-05	& &		&0.0097990	&1.16E-05	&   &		&0.0108990	&1.17E-05
\\ \hline
link 1	&0.0121072	&1.08E-04	&0.0111038	&1.01E-04	&0.0089001	&1.41E-05	&0.0100902	&3.09E-05	&0.0103592	&7.05E-06	&0.0102649	&7.62E-06
 \\ \hline
link 2	&0.0090666	&7.00E-05	&0.0106143	&9.76E-05	&0.0106210	&2.17E-05	&0.0091914	&1.12E-05	&0.0097514	&8.37E-06	&0.0111241	&1.65E-05
\\ \hline
link 3  &0.0070719	&4.21E-05	&0.0131189	&1.22E-04	&0.0089979	&1.90E-05	&0.0105010	&1.53E-05	&0.0104072	&1.17E-05	&0.0099058	&1.13E-05 \\ \hline
\end{tabular}
\caption{Estimate obtained by the tree estimator for 2 $\times$ 2 network, loss rates = $(1\%, 1\%) \times (1\%, 1\%, 1\%)$ }
\label{tree2-3}
\end{table*}
\begin{table*}[th]
  \centering
\scriptsize
\begin{tabular}{|c|r|r|r|r|r|r|} \hline
sample &\multicolumn{2}{|c|}{100 } & \multicolumn{2}{|c|}{500 } &\multicolumn{2}{|c|}{1000}	\\ \hline
Link & Mean & Var &	Mean & Var	& Mean & Var\\ \hline
source 1	&0.0089989	&4.90E-05	&0.01069910	&1.77E-05	&0.0103489	&1.50E-05 \\ \hline
source 2	&0.0094989	&7.48E-05	&0.00979914	&1.16E-05	&0.0108989	&1.17E-05 \\ \hline
link 1	&0.0116115	&5.37E-05	&0.00949612	&1.12E-05	&0.0103107	&3.92E-06 \\ \hline
link 2	&0.0098450	&3.44E-05	&0.00990625	&9.35E-06	&0.0104373	&4.20E-06 \\ \hline
link 3	&0.0100950	&4.34E-05	&0.00974978	&9.91E-06	&0.0101578	&4.45E-06 \\ \hline

\end{tabular}
\caption{Estimate obtained by the proposed estimators for 2 $\times$ 3 network, all loss rates =$(1\%, 1\%) \times (1\%, 1\%, 1\%)$ }
\label{proposed2-3}
\end{table*}

\section{Conclusion}

Network tomography has been around for a number of years and a large number of works have been done in the areas of loss tomography, delay tomography, topology tomography. Although most works reported in those areas focus on the tree topology, there are a few aiming at extending them to the general topology. Within the few, most believe there is a need to divide a general network into a number of trees despite there is a lack of a clear strategy how to divide a general network that can minimize information losses in estimation. We in this paper propose such a strategy that decompose a general network into a number of d-trees on the basis of intersections without information losses.
The proposed strategy consists of a divide-and-conquer strategy,  two estimators, and an estimation order.
The divide-and-conquer strategy  creates a consistent observation at the receivers attached to a d-tree. As a result, a likelihood function is constructed for a d-tree to connect the observations of the receivers to the unknown parameter (the pass rate of an intersection). From the likelihood function, we derive a MLE for the pass rate of a multi-source tree. The key differences between the MLE and those proposed for the tree topology are the parameter to be estimated and the number of probes considered in estimation. The proposed MLE considers all of the probes sent by the sources that makes it capable of supporting multiple sources. Apart from the estimator proposed for the multi-source tree, another estimator is proposed to estimate the loss rates of the upstream links of a multi-source tree.  Then, the  estimates obtained from a multi-source tree can be integrated into the estimates of the upstream links of that tree that make the estimates consistent.

Apart from the lack of estimators for the general topology, there has been a lack of statistical properties of an estimator regardless of the topology. Then.  experiments and simulation have been widely used to evaluate an estimator against another that lead to little reliable result since there are too many random factors affecting the results of experiments and simulations. To correct this, we in this paper derive
the statistical properties of the MLE and then the properties of the MVWA. With the help of the properties, we are able to evaluate some of the estimators proposed previously and reach some of the conclusion without running simulations or experiments.


To verify the points stated above, three sets
of simulations are conducted that show the estimates continuously improve in terms of mean and variance with the increase of the number of probes sent in an experiment.  The simulation results match the conclusions identified from the statistical properties.

\bibliography{../globcom06/congestion}
\end{document}